\documentclass[11pt]{article}
\usepackage{amsfonts,amssymb,a4}

\def\theequation{\thesection.\arabic{equation}}
\newcommand{\zeq}{\setcounter{equation}{0}}
\newcommand{\qed}{\hfill\rule{3mm}{3mm}}
\newtheorem{defi}{Definition}
\newtheorem{cor}{Corollary}
\newtheorem{lem}{Lemma}
\newtheorem{teo}{Theorem}

\newtheorem{conj}{Conjecture}
\newtheorem{pro}{Proposition}
\makeatletter \@addtoreset{equation}{section} \makeatother
\begin{document}


\voffset=-1.5truecm\hsize=16.5truecm    \vsize=24.truecm
\baselineskip=14pt plus0.1pt minus0.1pt \parindent=12pt
\lineskip=4pt\lineskiplimit=0.1pt      \parskip=0.1pt plus1pt

\def\ds{\displaystyle}\def\st{\scriptstyle}\def\sst{\scriptscriptstyle}


\let\a=\alpha \let\b=\beta \let\ch=\chi \let\d=\delta \let\e=\varepsilon
\let\f=\varphi \let\g=\gamma \let\h=\eta    \let\k=\kappa \let\l=\lambda
\let\m=\mu \let\n=\nu \let\o=\omega    \let\p=\pi \let\ph=\varphi
\let\r=\rho \let\s=\sigma \let\t=\tau \let\th=\vartheta
\let\y=\upsilon \let\x=\xi \let\z=\zeta
\let\D=\Delta \let\F=\Phi \let\G=\Gamma \let\L=\Lambda \let\Th=\Theta
\let\O=\Omega \let\P=\Pi \let\Ps=\Psi \let\Si=\Sigma \let\X=\Xi
\let\Y=\Upsilon



\def\\{\noindent}
\let\io=\infty

\def\VU{{\mathbb{V}}}
\def\ED{{\mathbb{E}}}
\def\GI{{\mathbb{G}}}
\def\Tt{{\mathbb{T}}}
\def\C{\mathbb{C}}
\def\LL{{\cal L}}
\def\RR{{\cal R}}
\def\SS{{\cal S}}
\def\NN{{\cal M}}
\def\MM{{\cal M}}
\def\HH{{\cal H}}
\def\GG{{\cal G}}
\def\PP{{\cal P}}
\def\AA{{\cal A}}
\def\BB{{\cal B}}
\def\FF{{\cal F}}
\def\TT{{\cal T}}
\def\v{\vskip.1cm}
\def\vv{\vskip.2cm}
\def\gt{{\tilde\g}}
\def\E{{\mathcal E} }
\def\I{{\rm I}}
\def\0{\emptyset}
\def\xx{{\V x}} \def\yy{{\bf y}} \def\kk{{\bf k}} \def\zz{{\bf z}}
\def\ba{\begin{array}}
\def\ea{\end{array}}  \def \eea {\end {eqnarray}}\def \bea {\begin {eqnarray}}

\def\tende#1{\vtop{\ialign{##\crcr\rightarrowfill\crcr
              \noalign{\kern-1pt\nointerlineskip}
              \hskip3.pt${\scriptstyle #1}$\hskip3.pt\crcr}}}
\def\otto{{\kern-1.truept\leftarrow\kern-5.truept\to\kern-1.truept}}
\def\arm{{}}
\font\bigfnt=cmbx10 scaled\magstep1

\newcommand{\card}[1]{\left|#1\right|}
\newcommand{\und}[1]{\underline{#1}}
\def\1{\rlap{\mbox{\small\rm 1}}\kern.15em 1}
\def\ind#1{\1_{\{#1\}}}
\def\bydef{:=}
\def\defby{=:}
\def\buildd#1#2{\mathrel{\mathop{\kern 0pt#1}\limits_{#2}}}
\def\card#1{\left|#1\right|}
\def\proof{\noindent{\bf Proof. }}
\def\qed{ \square}
\def\reff#1{(\ref{#1})}
\def\eee{{\rm e}}

\title{
On Lennard-Jones type potentials and hard-core potentials with an attractive tail
}
\author{
Thiago Morais$^{1,3}$, Aldo Procacci$^1$ and Benedetto Scoppola$^2$
\\
\small{$^1$ Departamento de Matem\'atica UFMG}
\small{ 30161-970 - Belo Horizonte - MG
Brazil}
\\
\small{$^2$ Dipartimento di Matematica
 Universit\'a ``Tor Vergata''}
\small{ V.le ricerca scientifica
00100 - Roma - Italy}
\\
\small{$^3$ Departamento de Matem\'atica UFOP}
\small{ 35400-000 - Ouro Preto - MG
Brazil}
}
\maketitle

\def\be{\begin{equation}}
\def\ee{\end{equation}}
\vskip.5cm

\begin{abstract}
We revisit  an old  tree graph  formula, namely the
Brydges-Federbush  tree identity,  and use it  to get  new bounds for the convergence
radius of the Mayer series for  gases  of continuous particles interacting via non absolutely summable  pair potentials
with   an attractive tail including  Lennard-Jones type pair potentials.

 \end{abstract}

\section{Introduction}

The rigorous approach  to the equilibrium statistical mechanics of
rarefied gases of classical particles is among the most  deeply studied subjects in
area of mathematical physics.  Most of the results in this research field  have
been obtained  during the
decade of the sixties. The rigorous analysis of the continuous gas of particles  has been mainly  developed by putting the system  in the Grand Canonical
Ensemble, in which there are three fixed thermodynamic parameters: the
 volume $V$ (i.e. the system is  confined in a large, typically
cubic, box), the  inverse temperature $\b$ and the   fugacity
$\l$. The total number of particles is not a fixed quantity. The  logarithm of the normalization constant of the
probability in such an  ensemble (i.e, the so-called  grand
canonical partition function), divided by the volume, is
proportional to the thermodynamic  pressure of the system, while its
derivative with respect to the fugacity $\l$ is proportional to the
density. Both pressure and  density are   functions of the two
thermodynamic parameters $\,\b$ and $\l$. So, reexpressing  the
fugacity as a function of the density, one can get the pressure as a
function of the temperature and density, i.e. the equation of state
of the system. It is thus crucial to be able to calculate the
logarithm of the grand canonical partition function in order to
study the thermodynamic properties of these systems.

The logarithm of the partition function can be written formally
in terms of a series in powers of the particle fugacity $\l$, known as the Mayer series, whose  coefficients (the Mayer coefficients) depend on the volume $V$
and on the inverse temperature $\b$.
Indeed, J. E. Mayer \cite{MM, May42} first gave the explicit expressions of the $n$-th order
coefficient of this series in terms of a sum  over connected graphs between $n$ vertices of cluster integrals.
 An upper bound
of the type $({\rm Const.})^n$ on these $n$-th order  coefficients, where $C$ is a constant (possibly depending on $\b$),
guarantees analyticity of the Mayer series, at least for sufficiently small activity values $\l$  (depending on the temperature but uniform in the volume).
\newpage Such a kind of bound
was generally  considered
very hard to obtain, due to the fact that the number $C_n$ of connected  graphs between $n$ vertices is greater
the order of $C^{n^2}$ with $C>1$ (indeed, by  a simple counting argument it is easy to show that $C_n\ge 2^{(n-1)(n-2)/2}$). So the question regarding the convergence
of  this series (and the related virial series) remained unanswered  until the beginning of the sixties.

The first breakthrough  towards the rigorous analysis  of the Mayer series for such systems was obtained in 1962 in a paper by
 Groeneveld \cite{gro62}, who gave, under the (severe) assumption that particles
interact via a purely repulsive pair potential, a bound of the type $({\rm Const.})^n$ for the  $n$-th order Mayer coefficient. Just one year  later, Penrose \cite{Pe1, Pe2} and independently  Ruelle \cite{Ru1,Ru2} proved that the Mayer series of a system of continuous particles
 is actually  an analytic function for small values of the fugacity for a large class of pair interactions (the so-called  stable and tempered pair potentials, see ahead for the definitions), as well as providing
a lower bound for the convergence radius, which remains till nowadays the best available in the literature.

These impressive results were  all obtained ``indirectly", i.e. not working directly on the  explicit expressions of the Mayer coefficients in terms of sums over connected graphs
by trying to bound them  exploiting some cancelations. The indirect
method used  was based on  the analysis of iterative relations between correlation functions of the system  (the so-called   Kirkwood-Salsburg Equations (KSE) \cite{Ki}).
A direct estimate on the Mayer coefficients was  proposed  some years later  by
Penrose \cite{pen67} who  however considered only systems of particles interacting via pair potentials in a quite restricted sub-class of the stable and tempered ones. Namely,    pair potentials
with a repulsive hard-core at short distance but possibly with a negative tail (i.e. attractive) at large distance. To get the ``direct" bound, Penrose  rewrote the sum over connected graphs
of the $n$-th order Mayer coefficient
in terms of trees, by grouping together some terms, obtaining, for the first time, a so-called {\it tree graph identity} (TGI). The method developed in \cite{pen67} was simpler than the
KSE technique used in \cite{Pe1, Pe2,Ru1,Ru2}. However,
the  lower bound on the convergence radius obtained by Penrose for  such restricted class of pair potentials was identical to that obtained
via KSE methods by Ruelle and himself in 1963. Probably for this reason this first example of TGI contained did not receive the attention it deserved.
The potentiality of the Penrose TGI
has been recently rescued in the recent works  \cite{FP,FP2,FPS} where the Penrose TGI has been used to get improvements the cluster expansion convergence region of
the abstract polymer gas \cite{FP}, the zero-temperature antiferromagnetic Potts model on infinite graphs \cite{FP2}, and of the hard-sphere gas on the continuum\cite{FPS}.

An alternative  TGI was proposed  a decade later in a paper
by Brydges and Federbush \cite{BF}. As far as absolutely summable potentials were considered, Brydges and Federbush were able to  deduce  new $C^n$ bounds on the $n$ order Mayer coefficient  (and hence on the Mayer series convergence radius).
These new bounds  improved those obtained by Penrose and Ruelle for a significant subclass the of absolutely  summable pair potentials.
Nevertheless the requirement  of absolute summability left out most of the physically relevant examples, such has
the hard-sphere gas and the Lennard-Jones gas  (both with non absolutely summable pair potentials due to their divergence at short distances).
The Brydges-Federbush TGI  had a much more successful
career, especially in constructive field theory,  and  further developments of this identity were  given  by several authors
(see e.g.  \cite{AR}, \cite{bry84}, \cite{BF},  \cite{pdls}). Anyway, the
limitations on the pair potentials present both  in the Penrose TGI and Brydges-Federbush TGI have substantially never been overcome and  the old bound obtained by Penrose and Ruelle in 1963 via
the KSE method remains until today the best  available,  valid moreover for the most general class of pair potentials, i.e. stable and tempered pair potentials.

A recent development of the  Brydges-Federbush TGI was given by one of us in \cite{Pr1, Pr}, within the framework of the abstract polymer gas. This development allowed to derive a new tree graph inequality
(see formula (3.11) in  \cite{Pr1}, or Proposition 1 in \cite{Pr}) which
was   used in \cite{Pr} to construct a cluster expansion for abstract polymers interacting via a non purely hardcore pair potential.
The results contained in \cite{Pr1, Pr} strongly indicate that
the range of application of the Brydges-Federbush TGI  could  be  broadened  even for continuous particle systems,  beyond the class of absolutely summable pair potentials.
Indeed, as pointed out by Poghosyan and Ueltschi  in \cite{PU}, the new inequality presented in \cite{Pr,Pr1} can be used  to get new bounds
for the class of potentials  considered by Penrose in 1967, e.g. short distance hard-core  potentials with an attractive tail (Theorem \ref{teo1} for  Penrose potentials ahead),  and these bounds  were explicitly  used in recent  works by Jansen \cite{J}   and Tate  \cite{T}.

In the present  paper we analyze the Brydges-Federbush tree graph identity under the light of the developments of \cite{Pr1, Pr}. We first revisit, for pedagogical purpose, the tree graph inequality obtained in \cite{Pr,Pr1} and illustrate  how from this inequality it is possible to obtain  straightforwardly the new bounds for
Penrose  potentials given in \cite{PU} (formula  (\ref{TGin}),   Theorem \ref{teo2} below). We then present  a new tree graph  inequality  (formula (\ref{TGru}), Theorem \ref{stabru}  below) which yields  alternative bounds for a wide class of  non-absolutely summable pair potentials with significant importance in physics. This class
includes the Lennard-Jones type potentials (see definition ahead) for which we are thus  able to produce new bounds alternative to the
classical Ruelle-Penrose bounds.   We also conjecture that this class actually coincides with the whole class of stable and tempered potentials.

While it easy to see that the inequality (\ref{TGin})  provides  a clearly improved bound with respect to the classical Penrose-Ruelle bound, as far as hard-core  potentials with an attractive tail are considered,
in the  case of the new inequality  (\ref{TGru}), applicable to Lennard-Jones type potentials, a direct comparison with the classical  bound appears to be quite involved and, in general, model-dependent, since it depends on the ability  to get an optimal  estimate for the stability constants  for given pair potentials. We are however able to produce  specific examples of
Lennard-Jones type pair potentials  for which our bound improves on the classical Ruelle-Penrose bound.  This
gives us  some hope   that the  new bounds  may lead  to an improvement on  the classical Penrose-Ruelle bound even for general  Lennard-Jones type pair potentials.
We plan to address such kind of  questions in a future paper.

\section{Notations and Results}
\zeq
\def\xx{{\bf x}}
\def\pp{{\bf p}}
Throughout the paper, if $S$ is a set, then $|S|$ denotes its cardinality.  If $n\in \mathbb{N}$ is a natural number then we will denote shortly $[n]=\{1,2,\dots,n\}$.
We also denote  $\mathbb{Z}^{+}=\mathbb{N}\cup\{0\}=\{0,1,2,\dots\}$.

\subsection{Continuous particle systems in the grand canonical ensemble}
We  will focus our attention on a system of  classical, identical
particles enclosed in a  cubic box $\L\subset \mathbb{R}^d$ with
volume $|\L|$. We denote by $x_i\in \mathbb{R}^d$ the position
$d$-dimensional vector  of the $i^{th}$ particle and by $|x_i|$ its modulus. We
assume that there are no particles outside $\L$ (free boundary
conditions) and that particles interact via a pair potential
$v(x_i,x_j)$ which, for sake of simplicity,  will be assumed
to be translational and rotational invariant. Namely we assume that
$$
v(x_i,x_j)\equiv V(|x_i-x_j|)
$$
with $V(r)$ being a function from $[0,+\infty)$ to $(-\infty,+\infty]$.
Given $N$
particles in  positions $(x_1,\dots ,x_N)\in \L^N$, their (configurational) energy $U(x_1,\dots, x_N)$ is given by
$$
U(x_1,\dots, x_N)=\sum_{1\le i<j\le N} V(|x_i-x_j|)
$$

\vv
\\{\bf Remark}. The value $V(r)=+\infty$ for some $r\ge 0$ is allowed.  In particular, given a pair potential
$V(|x|)$,
a pair $x_i, x_j$ is called {\it incompatible} if $V(|x_i-x_j|)=+\infty$, and we write  $x_i\nsim x_j$. Otherwise, if
 $V(|x_i-x_j|)<+\infty$, we say that $x_i$ and $x_j$ are {\it compatible} and we write $x_i\sim x_j$.

\begin{defi}\label{stability}
A pair potential $V(|x|)$ is said to be stable if there exists $B\ge 0$  such that,
for all $N\in \mathbb{N}$ and for all $(x_1,\dots, x_N)\in \mathbb{R}^{dN}$,
\be\label{2.6}
\sum_{1\le i<j\le N} V(|x_i-x_j|) \ge -B N
\ee
The $\inf$ of such $B$'s is called the {\it stability constant}.
\end{defi}

\begin{defi}\label{temperness}
A pair potential $V(|x|)$ is said to be tempered if
there exists a constant $r_0\ge 0$ such that
\be\label{cbetat}
\int_{|x|\ge r_0} |V(|x|)| dx <+\infty
\ee
\end{defi}

\begin{defi}\label{admi}
A  pair potential $V(|x|)$ is said to be admissible if it  is stable and tempered.
\end{defi}

\\Note that a pair potential $V(r)$ satisfying definition \ref{stability} is bounded below; namely,
by applying (\ref{2.6}) for the case $N=2$
$$
V(r)\ge -2B~~~~~~~~~~~~~~~~~~~~\forall r\ge 0
$$
Moreover, as a consequence of stability and  temperedness it is also easy to check that, for all $\beta>0$
\be\label{cbeta}
C(\beta)\doteq \int_{\mathbb{R}^d}
\left|e^{-\b V(|x|)} -1\right|
d x~<~+\infty
\ee
The grand canonical partition function $\Xi_\L(\b,\l)$ of the system is given by
\be\label{n2.2}
\Xi_\L(\b,\l)= 1+ |\L|\l+ \sum_{N\ge 2} {\l^N\over
N!}\int_{\L}dx_1\dots\int_{\L} dx_N
 e ^{-\b \sum_{1\le i<j\le N}V(|x_i-x_j|) }
\ee
with $\b>0$ being the inverse temperature, and $\l>0$ being the fugacity.
The pressure
of the system is  given by
\be\label{pressure}
P(\b,\l)=\lim_{|\L|\to \infty}{1\over\b |\L| }\log \Xi_\L(\b,\l)
\ee

\\The limit (\ref{pressure}) is known to exist if the pair potential $V(|x|)$ is admissible, i.e.  it is stable  and tempered (see e.g. \cite{Ru}, sections 3.3 and 3.4).
Moreover, a very well known and old result (see e.g. \cite{May42,MM, May47, Ru}) states that
 the factor $\log \Xi_\L(\b,\l)$ can be written in terms of a formal series in powers of $\l$. Namely,

\be\label{mayers}
{1\over |\L| }\log \Xi_\L(\b,\l)= \sum_{n\ge 1}
C_n(\b,\L){\l^n}
\ee
where $C_1(\b,\L)=1$ and, for $n\ge 2$,
\be\label{urse}
C_n(\b,\L)= {1\over |\L| }{1\over n!}\int_{\L}d\xx_1 \dots \int_{\L}
d\xx_n \sum\limits_{g\in G_{n}} \prod\limits_{\{i,j\}\in E_g}\left[
e^{ -\b V(|x_i -x_j|)} -1\right]
\ee
with   $G_n$ being the set of all connected graphs with vertex set
$[n]$ (here above $E_g$ denotes the edge-set of a graph $g\in G_n$).

\\The r.h.s. of (\ref{mayers}) is known as the Mayer series and the   term  $C_n(\b,\L)$ defined in (\ref{urse})
is the  $n$-th order Mayer coefficient (a.k.a. $n$-th order  connected cluster integral).
The dependence on the volume $|\L|$ of the Mayer coefficients $C_n(\b,\L)$ is, in  case of stable and tempered potentials, only marginal, and it is not difficult to show that
 $|C_n(\b,\L)|$  admits a  bound  uniform in $\L$ and, for every $n\in \mathbb{N}$,
the limit
\be\label{bninf}
C_n(\b)=\lim_{|\L|\to\infty}C_n(\b,\L)
\ee
exists and it is a finite constant (see e.g \cite{Ru}).
However, it is a completely different story  to  obtain an upper  bound on the (modulus of) $n$-order Mayer coefficient (\ref{urse})  with a good behavior
in $n$ (i.e. $C_n(\b)\sim [{\rm C}(\b)]^n$)  due to the fact that the cardinality of the set $G_n$ above is not less than $2^{(n-1)(n-2)/2}$. Such kind of bound must be obtained in general
 by exploiting   some  hidden cancellations  in the
factor
\be\label{fact}
\left|  \sum_{g\in G_{n}} \prod_{\{i,j\}\in E_g}\left[  e^{ -\b
V(|x_i -x_j|)} -1\right] \right|
\ee
It is clear that an efficient upper bound for the  Mayer coefficients $|C_n(\b,\L)|$, e.g. of the form $|C_n(\b,\L)|\le [{\rm C}(\b)]^n$ would immediately
yield a lower bound for the convergence radius of the
Mayer series of the pressure uniform in the volume.

As remarked in the introduction, as
far as  admissible potentials (i.e. satisfying condition (\ref{2.6})  and (\ref{cbetat})) are concerned,
the best rigorous bound on $|C_n(\b,\L)|$ available in the literature up to now was that
obtained by Penrose and Ruelle in 1963.
\begin{teo}[Penrose-Ruelle]\label{peru}
 Let $V(|x|)$ be an admissible pair  potential. Let $B$ be its stability constant. Then
the  $n$-order Mayer  coefficient $C_n(\b,\L)$ defined in (\ref{urse})
admits the bound
 \be\label{bmaru}
|C_n(\b,\L)|\le e^{2\b B (n-1)}n^{n-2} {[C(\b)]^{n-1}\over n!}
\ee
where  $C(\b) $ is the function defined in (\ref{cbeta}).

\\Consequently, the Mayer series in the r.h.s. of (\ref{mayers}) converges absolutely, uniformly in $\L$
 for any complex  $\l$ inside the disk
\be\label{radm}
|\l| <{1\over e^{2\b B+1} C(\b)}
\ee
\end{teo}

\\As said in the introduction, Brydges and Federbush gave an improvement of  the Penrose Ruelle bound as far as absolutely summable pair potentials are considered.

\begin{teo}[Brydges-Federbush]\label{teobf}
 Let $V(|x|)$ be an admissible and absolutely summable
 pair  potential. Let $B$ be its stability constant and let $\|V\|=\int_{\mathbb{R}^d}V(|x|)dx$. Then
the  $n$-order Mayer  coefficient $C_n(\b,\L)$ defined in (\ref{urse})
admits the bound
 \be\label{bmbf}
|C_n(\b,\L)|\le e^{\b B (n-1)}n^{n-2} {[\b \|V\|]^{n-1}\over n!}
\ee
\\Consequently, the Mayer series in the r.h.s. of (\ref{mayers}) converges absolutely, uniformly in $\L$
 for any complex  $\l$ inside the disk
\be\label{radbf}
|\l| <{1\over e^{\b B+1} \b \|V\|}
\ee
\end{teo}
It can be easily seen that, as far as absolutely integrable potentials are considered, (\ref{radbf}) may strongly improve (\ref{radm}). Let us consider for example
the so-called Morse  potential,
largely used in  simulations of   molecular gas models. The Morse potential is defined via the formula
$$
V_\r(|x|)= e^{2\r(1-|x|)}- 2 e^{\r(1-|x|)}
$$
where $\r>0$ is a constant. It is known that $V_\r(|x|)$ is stable for $\r\ge \ln 16$  (see \cite{Ru}, sec. 3.5.3, exercise  3B). Moreover $V_\r(|x|)$ is absolutely summable (it is bounded above by $e^{2\r}$ and bounded below by $-1$). Choosing $\b=1$ for simplicity and, e.g., $\r=6$,   an easy computation shows that for the Morse potential
$C(\b)\ge \int_{|x|\ge \ln2/6}|V_{\r=6}(|x|)|dx\ge 4\pi (182)$, while $\|V_{\r=6}\|\le 4\pi (204)$. So if $B_6$ is the stability constant of  $V_{\r=6}(|x|)$, the ratio between the Penrose-Ruelle bound (\ref{radm})  and the Brydges-Federbush   bound (\ref{radbf}) bound for the same convergence radius is smaller than $  (1.13) e ^{-B_6}$. To have an idea on how small is this ratio (so how much better is the Brydges-Federbush bound with respect to the
Penrose-Ruelle bound) for the Morse potential $V_{\r}(|x|)$ with $\r=6$ one can take a recent upper bound for $B_6=38.65$ given in \cite{AS}. Using this value we get that,  for inverse temperature $\b=1$,  the Brydges-Federbush bound  is at least $e^{38}$ larger than the Penrose-Ruelle bound!

Such a computation indicates  (or at least gives hope) that, being able to use the Brydges-Federbush  bound also for non absolutely integrable potentials, this could yield  improved bounds also in this case. Below we
exhibit a class of admissible non absolutely integrable potentials which have an unquestionable relevance in physical applications and  for which, as we will see below,  the Brydges-Federbush  TGI can be implemented.

\begin{defi}\label{ruelle}
An  admissible pair potential $V(|x|)$ on  $\mathbb{R}^d$ is said to be Ruelle  if
$$V(|x|)=\Phi_1(|x|)+\Phi_2(|x|)$$
where
$\Phi_1(|x|)$ is non-negative and tempered  and $\Phi_2(|x|)$ is stable and absolutely integrable (i.e such that  $\int_{\mathbb{R}^d}|\Phi_2(|x|)|dx <+\infty$).
\end{defi}
{\bf Remark}. Actually Ruelle  considered  potentials of the form  $V(|x|)=\Phi_1(|x|)+\Phi_2(|x|)$
but required  for $\Phi_2(|x|)$ to be,  more strongly, a {\it positive definite} pair potential
(i.e.  $\Phi_2(|x|)$ is a smooth function with positive Fourier transform). Indeed, any positive-definite potential $\Phi_2(|x|)$ is absolutely summable, because it admits Fourier transform, and
stable with stability constant $\Phi_2(0)/2$ (see \cite{Ru, ga}). Ruelle also conjectured that
any stable and tempered potential could always be written via such a decomposition, but this has be proven to be untrue in \cite{ANP, LS}.  However, the counterexamples  found in \cite{ANP, LS} are
absolutely summable stable potentials (so Ruelle according the definition \ref{ruelle}) and hence
it  seems reasonable  to conjecture that the class of Ruelle potentials introduced in Definition \ref{ruelle}, in which the condition for
$\Phi_2(|x|)$ to be  {\it positive-definite} is replaced by the much milder condition for $\Phi_2(|x|)$ to be  stable and absolutely summable, actually coincides with the whole class of admissible potentials.

\begin{conj}
Any stable and tempered pair potential can  always be written as a sum of a non-negative potential plus an absolutely summable stable potential.
\end{conj}

\\ The class of potentials introduced in definition \ref{ruelle} appears to be  in any case  very  big  and includes many
(if not all)  physically realistic pair potentials. In particular it  contains the so-called  Lennard-Jones type potentials.
\begin{defi}\label{LenJ}
An  admissible pair potential $V(|x|)$ on  $\mathbb{R}^d$ is said to be of  Lennard-Jones type  if there exist constants $w$,
$r_1$, $r_2$  (with $r_1\le r_2$)  and  non-negative monotonic decreasing functions
$\xi(|x|)$, $\eta(|x|)$ such that

\be\label{condLJ}
V(|x|)\cases{ \ge \xi(|x|) & if $|x|\le r_1$\cr\cr
\ge - w &  if $r_1\le |x|\le r_2$\cr\cr
\ge -\eta(|x|) & if $|x|\ge r_2$
}
\ee
with
\be\label{short}
 \int_{|x|\le r_1} \xi (|x|) dx =+\infty
\ee
and
\be\label{long}
\int_{|x|\ge r_2} \eta (|x|) dx <+\infty
\ee
\end{defi}
Indeed, in the appendix of the present paper we will  prove the following  proposition.
\begin{pro}\label{lj}
Let $V(|x|)$ a Lennard-Jones type pair potential according to Definition \ref{LenJ}.
Then $V(|x|)$  is Ruelle according to definition \ref{ruelle}.
\end{pro}

\\An interesting subclass of the pair potentials satisfying Definition \ref{ruelle}  is the   class  of hard-core potentials with an attractive tail originally considered  by Penrose in \cite{pen67}.

\begin{defi}[Penrose pair potential]\label{hard2a}
An admissible pair potential $V(|x|)$ on $ \mathbb{R}^d$ is called Penrose
if it has an hard-core, i.e.  if there exists a positive constant $a>0$ such that $V(|x|)=+\infty$ whenever $|x|\le a$ and $V(|x|)<0$ whenever $|x|> a$.
The constant $a$ is called the hard-core radius of the Penrose pair potential.
\end{defi}

\\As mentioned in the introduction, Poghosyan and Ueltschi pointed out in \cite{PU} that for Penrose potentials the  inequality (\ref{TGin}) proposed in \cite{Pr,Pr1} yields  new
bounds  which improve the classical result of Theorem \ref{peru}.

\begin{teo}\label{teo1} Let $V(|x|)$ be a Penrose potential according to Definition \ref{hard2a}. Let $B$ be its stability constant and let $a$ be its hard core radius.
Then $n$-order Mayer  coefficient $C_n(\b,\L)$ defined in (\ref{urse})
admits the bound
\be\label{bo1}
|C_n(\b,\L)|\le e^{\b B n}n^{n-2} {[C^*(\b)]^{n-1}\over n!}
\ee
where, if we denote by $W_a(d)$ the volume of the sphere of radius $a$ in $d$ dimensions,
\be\label{cstarb}
C^*(\b) = W_a(d) + \b\int_{|x|\ge a} |V(|x|)| dx
\ee
Consequently, the Mayer series converges absolutely for all complex activities $\l$  such  that
\be\label{radmpen}
|\l|<{1\over e^{\b B+1} C^*(\b)}
\ee
\end{teo}

\\{\bf Remark}. Formula (\ref{radmpen}) represents an improvement on the classical bound (\ref{radm})
for all Penrose potentials with an attractive tail, i.e. those $V$ for which $V(|x|)<0$ when $|x|> a$ (which are, as a matter of fact, the  interesting cases).
Indeed, for such $V$, in formula (\ref{radmpen})
 the presence of the factor
$e^{\b B}$  improves (exponentially) the factor $e^{2\b B}$ of formula (\ref{radm}). Moreover
 $C^*(\b) < C(\beta)$ for all $\b$ since $e^x-1> x$ for all $x>0$. We will  give  below, for the benefit of the reader, a self-contained  proof  of  Theorem \ref{teo1}.

\subsection{Results}
\\The main result of the paper can be summarized by the following   theorem.

\begin{teo}\label{teo2}
Let $V(|x|)=\Phi_1(|x|)+\Phi_2(|x|)$ be Ruelle  potential in the sense of Definition \ref{ruelle} and let  $\tilde B$ the stability constant of the potential $\Phi_2(|x|)$.
 Then the $n$-th order Mayer coefficient $C_n(\b,\L)$ defined in (\ref{urse})
admits the bound
\be\label{bteo2}
|C_n(\b,\L)|\le e^{{\b \tilde B}n}~n^{n-2} {[\tilde C(\b)]^{n-1}\over n!}
\ee
where
\be\label{tilb}
\tilde C(\b) =\int \left[ |e ^{-\b \Phi_1(|x|)} -1|+ \b|\Phi_2(|x|)|\right]dx
\ee
Consequently, the Mayer series converges absolutely for all complex activities $\l$  such  that
\be\label{radmru}
|\l|<{1\over e^{{\b \tilde B}+1} \tilde C(\b)}
\ee
\end{teo}

\\{\bf Remark}.  If the Ruelle potential $V(|x|)$ is absolutely integrable, e.g. $V(|x|)=\Phi_2(|x|)$, then (\ref{radmru}) is the Brydges-Federbush bound (\ref{radbf}).
On the other hand, if  $V(|x|)$ is not absolutely integrable then (\ref{radmru}) is a new bound.
 In  this case however a comparison of bound (\ref{radmru}) with the original Ruelle-Penrose bound (\ref{radm}) appears to be more tricky,
since the possible improvement of formula (\ref{radmru}) with respect to
the classical (\ref{radm})  is strongly  model dependent and, for a
fixed potential, could rely on the search of an optimal
decomposition $V(|x|)=\Phi_1(|x|)+\Phi_2(|x|)$ providing the as small
as possible constant $\tilde B$ and, at the same time, the as small as
possible quantity $\tilde C(\beta)$. The sharp evaluation of the
stability  constant of a given  potential is quite a hard subject,
and it is beyond the scope of this paper. However, it is not difficult to  exhibit
examples of  non absolutely integrable pair potentials (actually the example below is a
Lennard-Jones type potential)  for which bound (\ref{tilb}) strongly
improves on (\ref{radm}). E.g. an  example goes as follows. Let
$\Phi_2(|x|)$ is an absolutely summable stable potential  for which
 the minimum particle distance $r_{\rm min}$ in a minimal energy configuration is strictly positive
(such kind of potentials do exist: e.g. Morse potentials have  $r_{\rm min}>0$, see \cite{LoS}). Let $\tilde B$ the stability constant of $\Phi_2(|x|)$.
Let now $\x(r)$ be a positive function such that
$\xi(|x|)> \Phi_2(|x|)$ for $|x|< r_{\rm min}$  and $\int_{|x|<r_{\rm min}}\x(|x|)dx=\infty$. Let
$$
\Phi_1(|x|)=\cases{
\xi(|x|)- \Phi_2(|x|) &if $|x|< r_{\rm min}$\cr 0 & if $|x|\ge r_{\rm min}$
}
$$
Then  $V(|x|)=\Phi_1(|x|)+\Phi_2(|x|)$  is a Lennard-Jones type potential given by
$$
V(|x|)=\cases{
\xi(|x|) &if $|x|< r_{\rm min}$\cr \Phi_2(|x|) & if $|x|\ge r_{\rm min}$
}
$$
By construction, $V(x)$ has the same stability constant $\tilde B$ as $\Phi_2(|x|)$.  Indeed,  minimal energy configurations for $V(|x|)$ are also minimal energy configurations
for $\Phi_2(|x|)$ (since $V(|x|)=\Phi_2(|x|$ for $|x|\ge r_{\rm min}$) and any energy configuration for $V(|x|)$ in which there are pairs of particles at distance
less than $r_{\rm min}$ has larger energy than
the same configuration for $\Phi_2(|x|)$ (because $V(|x|)>\Phi_2(|x|)$ for $|x|< r_{\rm min}$). Proceeding now as we did in the Morse potential example after Theorem
\ref{teobf}, we get that the ratio between
the Penrose-Ruelle bound and Brydges-Federbush bound for  such $V(|x|)$ with $\b=1$ is  $e^{-\tilde B} (C_1/C_2)$  where $C_1=\int_{\mathbb{R}^d}|e^{-V(|x|)} -1|dx$
and $C_1=\int_{\mathbb{R}^d}|\Phi_2(|x|)dx$.
Picking $\Phi_2$ and $\xi$ in such a way that   $C_1/C_2 = O(1)$  we get, similarly  to the case presented  in the remark after Theorem \ref{teobf}, that the  bound (\ref{radmru})
is  $O(1)\exp\tilde B$ times larger than the  Penrose-Ruelle bound.

\vv
\\The rest of the paper is organized as follows.
In  section 3 we  will  revisit  the  Brydges-Federbush TGI, introducing the necessary notations, and we will state and prove two technical results, namely Theorem \ref{stabbry} and
Theorem \ref{stabru}, which will be the main tools in order to prove Theorems \ref{teo1} and \ref{teo2} respectively, whose proofs will be completed in section 4.
Finally in the appendix we will   prove Proposition \ref{lj}.

\section{Algebraic Brydges-Federbush tree graph identity}\label{algebr}
\zeq
\subsection{Pair potentials in $[n]$}
In this section we present the Brydges-Federbush  identity which in turn leads to alternative expressions for
the Mayer  coefficients
defined in (\ref{urse}). This alternative expression of the Mayer coefficients
is  written in terms of a sum over trees rather than connected graphs and, as we will see,  this fact permits us
to get rid of the combinatorial problem. The Brydges-Federbush  identity is  essentially algebraic
and in order to introduce it  we need to give some preliminary notations.
We recall that we let $[n]=\{1,2,\dots,n\}$. We also denote by ${\rm E}_n$ the set of all unordered pairs in $[n]$ and, in general, if $X\subset [n]$, then $E_X$ will denote
the set of all unordered pairs in $X$.
\def\E{{\rm E}}
\begin{defi}\label{pair}
An algebraic  pair interaction in   $[n]$ is a map $V:
\E_n\rightarrow \mathbb{R}\cup\{+\infty\}$ that associate to any
unordered pair in $\{i,j\}\in E_n$ a number $V_{ij}$  (with the convention, $i<j$) with values in $ \mathbb{R}\cup\{+\infty\}$

\end{defi}
\def\N{\mathbb{N}}
\begin{defi}\label{hard}
Let $V$ be an algebraic  pair interaction in  $[n]$. Let $\{i,j\}\in \E_n$. If $V_{ij}=+\infty$  we say that the pair
$\{i,j\}$ is incompatible and  we write $i\not\sim j$. Otherwise,
if $V_{ij}<+\infty$  we say that the pair
$\{i,j\}$ is compatible and  we write $i\sim j$. A set $X\subset [n]$ is called incompatible if
there are $\{i,j\}\in E_X$ such that $i\not\sim j$. Otherwise, if for every $\{i,j\}\in E_X$ we have that $i\sim j$
then $X$ is called a compatible set.
\end{defi}


\begin{defi}\label{staby} An algebraic  pair interaction $V$ in $[n]$
is said to be stable if there exists a constant $B\ge 0$
 such that, for all
$X\subset [n]$ with  $|X|\ge 2$  we have
\be\label{stabc1}
\sum_{\{i,j\}\in E_X}V_{ij}\geq -B|X|
\ee
\end{defi}

\\Observe that, if  $V(|x|)$ is a stable  pair potential in $\mathbb{R}^d$ in the sense of definition
\ref{stability} with stability constant (not greater than) $B$, then  for any $n \in \mathbb{N}$ and any
$n$-tuple $(x_1,\dots,x_n)\in \mathbb{R}^{dn}$, the algebraic pair interaction  $V_{ij}= V(|x_i-x_j|)$ has  stability constant (not greater than) $B$ in
the sense of definition \ref{staby}.
Note moreover that we can restrict ourselves  to check the condition  (\ref{stabc1})
only for those $X$ which do not contain incompatible pairs, otherwise the l.h.s. of (\ref{stabc1})
takes the value $+\infty$.

\begin{defi}\label{bounded}
An algebraic  pair interaction $V$ in $[n]$ is
said to be
 {\it repulsive}  if, $\forall\{i,j\}$
$$
V_{ij}\geq 0
$$
and $V$  is said to be
 {\it bounded}  if, $\forall\{i,j\}$
$$
V_{ij}< +\infty
$$
\end{defi}

\begin{defi}\label{staru}
An algebraic  pair interaction $V$ in $[n]$ is said to be Ruelle-stable  if $V=\Phi_1+\Phi_2$ where
$\Phi_1$ is a repulsive pair interaction in $[n]$ and $\Phi_2$ is a bounded and stable   pair interaction in $[n]$.
\end{defi}

\\Of course,  if $V(|x|)= \Phi_1(|x|)+ \Phi_2(|x|)$ is a Ruelle pair potential in $\mathbb{R}^d$,
then for any $n\in \mathbb{N}$ and any $n$-tuple $(x_1,\dots, x_n)\in \mathbb{R}^{dn}$ the algebraic pair potential $V_{ij}$
in $[n]$ defined by $V_{ij}= V(|x_i-x_j|)$ is Ruelle-stable.

\subsection{The Brydges-Federbush tree graph identity}
We  are now ready to state the (algebraic) Brydges-Federbush tree graph identity.
Given an algebraic pair interaction $V$ in $[n]$, consider the factor
\be\label{2.4s}
\sum_{g\in {G}_{n}}\prod_{\{i,j\}\in
E_g}\left(e^{- V_{ij}}-1\right)
\ee
where $G_n$ is the set of connected graph in $[n]$ and, for $g\in G_n$, $E_g$ is the edge set of $g$. Then the following theorem holds.

\begin{teo}\label{teoTGI}
Let  $V$ be a bounded algebraic pair interaction  in $[n]$,
then the following identity holds:

\be\label{TGI}
\sum_{g\in G_n} \prod_{\{i,j\}\in E_g}\left(e^{-  V_{ij}}-1\right)~=~
\sum_{\t\in T_n}\prod_{\{i,j\}\in E_\t} (- V_{ij}) \int
d\m_{\t}({\bf t}_{n},{\bf X}_{n}) e^{-  \sum_{1\leq i<j\leq n}
{\bf t}_{n} (\{i,j\})V_{ij}}
\ee

\\where

\begin{itemize}
\item[-] $T_n$ denotes the set of all trees with vertex set $[n]$ and, for $\t\in T_n$, $E_\t$ is the edge set of $\t$;

\item[-] ${\bf t}_{n}$ denotes a set on $n-1$ interpolating parameters
${\bf t}_{n}\equiv (t_1 ,\dots ,t_{n-1})\in [0,1]^{n-1}$;

\item[-] the
symbol ${\bf X}_{n}$ denotes a set of ``increasing'' sequences of
$n-1$ subsets, ${\bf X}_{n}\equiv X_1 ,\dots ,X_{n-1}$ such that
$\forall i\in [n-1]$,  $X_{i}\subset [n]$,
$X_{i}\subset X_{i+1}$, $|X_{i}|~=~i$ and $X_1 ~=~\{1\}$.

\item[-] The factor ${\bf t}_{n} (\{i,j\})$, which depends on ${\bf X}_n$, is defined as
$$
{\bf t}_{n} (\{i,j\}) = t_{1}(\{i,j\})\dots t_{n-1}(\{i,j\})
$$
with, for $s\in[n-1]$ and $\{i,j\}\in E_n$,
$$
t_{s}(\{i,j\})~=~\cases{ t_{s}\in [0,1] &if $i\in X_s$ and $j\notin
X_s$ or viceversa\cr 1 &otherwise\cr}
$$

\item[-]The measure $\m_{\t}({\bf t}_{n},{\bf X}_{n})$ is the following probability measure

$$
\int  d\m_{\t}({\bf t}_{n},{\bf X}_{n})[...] ~=~
\int_{0}^{1}dt_1\dots\int_{0}^{1}dt_{n-1} \sum_{{\bf X}_{n}\atop
{\rm compatible~with}\ \t} t_{1}^{b_{1}-1}\dots
t_{n-1}^{b_{n-1}-1} [...]
$$

\\where
${\bf X}_{n}=(X_1,\dots, X_{n-1})$ compatible with $\t$ means that $\forall i\in [n-1]$, $X_i$ contains exactly $i-1$ edges of $\t$ and
$b_i$ is the number  of edges  of $\t$ which have one vertex in $X_i$ and the other one in $[n]\setminus X_i$.

\end{itemize}

\end{teo}
We refer the reader to references \cite{bry84}  and \cite{pdls}   for a detailed  proof.
\vv\vv
\\\\{\bf Remark}. The hypothesis that $V$ has to be a bounded pair interaction is necessary to give
meaning to the r.h.s. of (\ref{TGI}), in particular to the factor $\prod_{\{i,j\}\in \t} (- V_{ij}) $.
However Theorem \ref{teoTGI} immediately implies the following corollary.

\begin{cor}\label{TGIc}
Let  $V$ be a pair interaction  in $[n]$ not necessarily bounded (i.e.  $V_{ij}=+\infty$ for some $\{i,j\}\in \E_n$ is allowed).
Then  following identity holds:

\be\label{TGI2}
\sum_{g\in G_n} \prod_{\{i,j\}\in g}\left(e^{-  V_{ij}}-1\right)~=~\lim_{H\to +\infty}
\sum_{\t\in T_n}\prod_{\{i,j\}\in \t} (-V^H_{ij}) \int
d\m_{\t}({\bf t}_{n},{\bf X}_{n}) e^{-   \sum_{1\leq i<j\leq n}
{\bf t}_{n} (\{i,j\})V^H_{ij}}
\ee

where $V^H$ is the bounded pair potential on $[n]$ given by
\be\label{vuha}
V^H_{ij}=\cases{ H & if $V_{ij}=+\infty$\cr\cr
V_{ij} & otherwise
}
\ee
\end{cor}
\\{\bf Proof}. Indeed, by theorem \ref{TGI} we have that, for all $H\in \mathbb{R}$,
$$
\sum_{\t\in T_n}\prod_{\{i,j\}\in E_\t} (-V^H_{ij}) \int
d\m_{\t}({\bf t}_{n},{\bf X}_{n}) e^{-  \sum_{1\leq i<j\leq n}
{\bf t}_{n} (\{i,j\})V^H_{ij}}= \sum_{g\in G_n} \prod_{\{i,j\}\in E_g}\left(e^{-  V^H_{ij}}-1\right)
$$
and
$$
\sum_{g\in G_n} \prod_{\{i,j\}\in E_g}\left(e^{-  V_{ij}}-1\right)~=~\lim_{H\to +\infty}
\sum_{g\in G_n} \prod_{\{i,j\}\in E_g}\left(e^{- V^H_{ij}}-1\right)
$$
$\Box$
\vv

\\We finally conclude this section  by recalling two  lemmas   which will be used below. We refer again the reader to references
\cite{bry84} and \cite{pdls} for their proofs.

\begin{lem}\label{posit}
Let $V$ be a bounded  algebraic pair potential in $[n]$,  then for any  $\t\in T_n$
it holds the identity

\be\label{TGI3}
\prod_{\{i,j\}\in E_\t}|e^{-  V_{ij}}-1|\,=\,
\prod_{\{i,j\}\in E_\t} |V_{ij}|
\int d\m_{\t}({\bf t}_{n},{\bf X}_{n})
e^{- \sum_{\{i,j\}\in E_\t}
{\bf t}_{n}(\{i,j\})V_{ij}}
\ee

\end{lem}

\begin{lem}\label{convex}
Let $V_{ij}$  a stable  algebraic   pair interaction in $[n]$ with stability constant $B$, then, for all  $X_1 ,\dots X_{n-1}$ and
all $(t_1 ,t_2 ,\dots t_{n-1})\in [0,1]^{n-1}$

$$
\sum_{1\leq i<j\leq n}
{\bf t}_{n} (\{i,j\})V_{ij} \geq -n B
$$
\end{lem}

\subsection{New Tree graph inequalities from Brydges-Federbush TGI}

We are now ready to enunciate the two main technical results derived from the algebraic Brydges-Federbush tree graph identity. The first of them
(Theorem \ref{stabbry} below)  has been originally proved in \cite{Pr1,Pr}. The second result (Theorem \ref{stabru} below) is, as far as we know, a new result.
Theorem \ref{stabbry} will be used later to obtain improved bounds
 (respect to the classical bounds (\ref{bmaru})-(\ref{radm})) for
the convergence radius of the Mayer expansion in systems of particles interacting through a Penrose pair potential according to Definition \ref{hard2a}. Theorem \ref{stabru} will be used later to obtain new bounds,
alternative to the classical ones (\ref{bmaru}) and  (\ref{radm}), in systems of particles interacting through a Ruelle pair potential according to Definition \ref{ruelle}.

\begin{teo}\label{stabbry}
Let $V$ be a stable algebraic pair interaction in $[n]$ with stability constant $B$, then the following
inequality holds:

\be\label{TGin}
\bigg|\sum_{g\in G_n}\prod_{\{i,j\}\in E_g}
\left(e^{- V_{ij}}-1\right)\bigg|\leq
e^{n  B}\sum_{\t\in T_n}\prod_{\{i,j\}\in E_\t}
F_{ij}
\ee
where
\be\label{fij}
F_{ij}=
\cases{ |e^{- V_{ij}}-1|\equiv 1 ~~~& if ~$V_{ij}=+\infty$ \cr\cr
 |V_{ij}| ~~~& otherwise
}
\ee
\end{teo}

\\{\bf Proof}. By theorem \ref{TGI} and corollary \ref{TGIc} we have that
$$
\Big|\sum_{g\in G_n} \prod_{\{i,j\}\in E_g}\left(e^{-V_{ij}}-1\right)\Big|=
\lim_{H\to\infty }
\Big|\sum_{g\in G_n} \prod_{\{i,j\}\in E_g}\left(e^{-  V^H_{ij}}-1\right)\Big|=
$$
$$
=\lim_{H\to +\infty}\Big|
\sum_{\t\in T_n}\prod_{\{i,j\}\in E_\t} (-V^H_{ij}) \int
d\m_{\t}({\bf t}_{n},{\bf X}_{n}) e^{- \sum_{1\leq i<j\leq n}
{\bf t}_{n} (\{i,j\})V^H_{ij}}\Big|\le
$$
$$
\le \lim_{H\to +\infty}
\sum_{\t\in T_n}\prod_{\{i,j\}\in E_\t} |V^H_{ij}| \int
d\m_{\t}({\bf t}_{n},{\bf X}_{n}) e^{-  \sum_{1\leq i<j\leq n}
{\bf t}_{n} (\{i,j\})V^H_{ij}}\doteq \lim_{H\to +\infty}
\sum_{\t\in T_n}w^H_\t
$$
where\def\E{{\rm E}}
$$
 w^\t_H=
\prod_{\{i,j\}\in E_\t} |V^H_{ij}|
\int d\m_{\t}({\bf t}_{n},{\bf X}_{n})
e^{-  \sum_{1\leq i<j\leq n}{\bf t}_{n} (\{i,j\})V^H_{ij}}
$$
Observe now that, for any tree $\t\in T_n$,  the edge set { $E_\t$
is naturally partitioned}  into two disjoint sets
$E_\t^H$ and $E_\t\backslash E_\t^H$ where
$$
E_\t^H\,\,= \,\,\{\{i,j\}\subset E_\t : i\not\sim j\}
$$
So, by definition (\ref{vuha})
$$
w^\t_H
= H^{|E_\t^H|}\prod_{\{i,j\}\in E_\t\backslash E_\t^H}  |V_{ij}|
{\int d\m_{\t}({\bf t}_{n},{\bf X}_{n})                                                                                                                          }{}
e^{{ -  \sum_{1\leq i<j\leq n}
{\bf t}_{n} (\{i,j\})V^H_{ij}}}
$$
Now, we can rewrite

$$
\sum_{1\leq i<j\leq n}
{\bf t}_{n}(\{i,j\})V^H_{ij}
~=~
\sum_{1\leq i<j\leq n}
{\bf t}_{n}(\{i,j\})U^{(1-\e)H}_{ij}
~+~\sum_{1\leq i<j\leq n}
{\bf t}_{n}(\{i,j\})V^{\e H}_{ij}
$$
where $\e>0$ and
$$
U^{(1-\e)H}_{ij}=\cases{ (1-\e)H & if $i\nsim j$\cr\cr
0 &otherwise
}
$$
and
$$
V^{\e H}_{ij}=\cases{ \e H & if $i\nsim j$\cr\cr
V_{ij} &otherwise
}
$$
\\The algebraic pair interaction $V^{\e H}$ in $[n]$,
when  $H$ is taken sufficiently large, is stable with the stability constant equal to that of  the algebraic pair interaction $V$ in the sense of definition
\ref{staby}. Namely, there exists an $H_0>0$ (which depends on $V$ and $n$)
such that for all  $H\ge H_0$

$$
\sum_ {\{i,j\}\in E_X}V^{\e H}_{ij}\ge -B|X|
$$
for all $X\subset [n]$. So, by proposition \ref{convex} we have

$$
\sum_{1\leq i<j\leq n}
{\bf t}_{n}(\{i,j\})V^{\e H}_{ij}\ge
-nB
$$
and thus we can bound, for $H\ge H_0$
\be\label{bouf}
 w^\t_H\,\,\le\,\,e^{ B n} \Big[\prod_{\{i,j\}\in E_\t\backslash E_\t^H}  |V_{ij}|\Big]
H^{|E_\t^H|}
\int d\m_{\t}({\bf t}_{n},{\bf X}_{n})
e^{- \sum_{1\leq i<j\leq n}
{\bf t}_{n}(\{i,j\})U^{(1-\e)H}_{ij}}
\ee

\\On the other hand the potential $U^{(1-\e)H}_{ij}$ is non negative, so

$$
\sum_{1\leq i<j\leq n}
{\bf t}_{n}(\{i,j\})U^{(1-\e)H}_{ij}
\ge
\sum_{\{i,j\}\subset E_\t^H}
{\bf t}_{n}(\{i,j\})U^{(1-\e)H}_{ij}=
$$
$$
=
\sum_{\{i,j\}\subset E_\t^H}
{\bf t}_{n}(\{i,j\})(1-\e)H
+
\h\sum_{\{i,j\}\subset E_\t\backslash E_\t^H}
{\bf t}_{n}(\{i,j\}) -\h
\sum_{\{i,j\}\subset E_\t\backslash E_\t^H}
{\bf t}_{n}(\{i,j\})~~\ge
$$
$$
\ge
\sum_{\{i,j\}\subset E_\t^H}
{\bf t}_{n}(\{i,j\})(1-\e)H+\!\!\!\!\!
\sum_{\{i,j\}\subset E_\t\backslash E_\t^H}
{\bf t}_{n}(\{i,j\}))\h  - {|E_\t\backslash E_\t^H|\h}
$$
\be\label{edue}
\ge
\sum_{\{i,j\}\subset E_\t}
{\bf t}_{n}(\{i,j\})V^\t_{ij} - \eta|E_\t\backslash E_\t^H|
\ee

\\where $V^\t_{ij}$ is  the positive ($H$ dependent) pair potential given by
\be\label{defvt}
V^\t_{ij}=\cases{(1-\e)H &if $\{i,j\}\in E_\t^H$ \cr\cr
{\h} & if $\{i,j\}\in E_\t\backslash E_\t^H$}
\ee

\\Plugging (\ref{defvt}) into   (\ref{edue}) and observing that, by definition (\ref{defvt}), we can write
$$
H^{|E_\t^H|}=
\left[{1\over \h}\right]^{|E_\t\backslash E_\t^H|}
\left[{1\over 1-\e}\right]^{|E_\t^H|}
\prod_{\{i,j\}\in  E_\t} V^\t_{ij}
$$

\\we arrive at
$$
w^\t_H\le
e^{ n B}\,
\left[\prod_{\{i,j\}\in E_\t\backslash E_\t^H} |V_{ij}|\right]\times
$$
$$
\times\left[{e^\h\over \h}\right]^{|E_\t\backslash E_\t^H|}\left[{1\over 1-\e}\right]^{|E_\t^H|}
\times
{ \prod_{\{i,j\}\in  E_\t}  V^\t_{ij}
 \int d\m_{\t}({\bf t}_{n},{\bf X}_{n})
e^{-  \sum_{\{i,j\}\subset E_\t}
{\bf t}_{n}(\{i,j\})V^\t_{ij} }}
$$

\\Using now Lemma \ref{posit}, i.e. formula (\ref{TGI3}) we have that
$$
{\prod_{\{i,j\}\in  E_\t}  V^\t_{ij}
 \int d\m_{\t}({\bf t}_{n},{\bf X}_{n})
e^{-  \sum_{\{i,j\}\subset E_\t}
{\bf t}_{n}(\{i,j\})V^\t_{ij} }}=
\prod_{\{i,j\}\in E_\t}\left|e^{-V^\t_{ij}}-1\right|=
$$
$$
=
 \prod_{\{i,j\}\in E_\t^H}\left|e^{-U^{(1-\e)H}_{ij}}-1\right|
\left|e^{-\h}-1\right|^{|E_\t\backslash E_\t^H|}\le
 \prod_{\{i,j\}\in E_\t^H}\left|e^{-V_{ij}}-1\right|
\left|e^{-\h}-1\right|^{|E_\t\backslash E_\t^H|}
$$
where in the last line we have used that
$U^{(1-\e)H}_{ij}< V_{ij}$ for all $H>0$ and for all $\{i,j\}\subset [n]$ such that $i\nsim j$.
Thus we get, for $H\ge H_0$
$$
w^\t_H\le
e^{ n B}
\left[\prod_{\{i,j\}\in E_\t\backslash E_\t^H} |V_{ij}|\right]\times
$$
$$
\times\left[{e^\h\left|e^{-\h}-1\right|\over \h}\right]^{|E_\t\backslash E_\t^H|}\left[{1\over 1-\e}\right]^{|E_\t^H|}
\times
 \prod_{\{i,j\}\in E_\t^H}\left|e^{-V_{ij}}-1\right|=
$$
$$
=e^{ n B}
\left[\prod_{\{i,j\}\in E_\t\backslash E_\t^H} {(e^\h-1)\over \h}|V_{ij}|\right]
 \prod_{\{i,j\}\in E_\t^H}{1\over 1-\e}\left|e^{-V_{ij}}-1\right|=
$$

\\Hence, since $\h$  and $\e$ can be taken as small as we please, we get finally

$$
w^\t_H\,\,\le e^{ nB}
\prod_{\{i,j\}\in E_\t}  F_{ij}
$$
where $F_{ij}$ is precisely the function defined in (\ref{fij}). $\Box$

\begin{teo}\label{stabru}
Let $V=\Phi^1+\Phi^2$ be a Ruelle algebraic pair interaction  in $[n]$ (in the sense of  Definition \ref{ruelle})  with $\Phi^1$ non-negative
and $\Phi^2$ stable. Let $B_0$  the stability constants associated to $\Phi_2$.
 Then the following
inequality holds:

\be\label{TGru}
\bigg|\sum_{g\in G_n}\prod_{\{i,j\}\in E_g}
\left(e^{-V_{ij}}-1\right)\bigg|\leq
e^{n B_0}\sum_{\t\in T_n}\prod_{\{i,j\}\in E_\t}
\Big[|e^{-\Phi^1_{ij}}-1|+ |\Phi^2_{ij}|\Big]
\ee
\end{teo}

\vv
\\{\bf Proof}.
First observe that by definition $\Phi^2$ is a bounded potential, so if for some $\{i,j\}$ we have that $V_{ij}=+\infty$ then $\Phi^1_{ij}=+\infty$
while $\Phi^2_{ij}<+\infty$. Hence we define
$$
\Phi^{1,H}_{ij}= \cases{ H & if $\Phi^1_{ij}=+\infty$\cr\cr
\Phi^1_{ij} & otherwise
}
$$
and
$$
V_{ij}^H= \Phi^{1,H}_{ij}+\Phi^{2}_{ij}
$$

\\Then by theorem \ref{teoTGI} and corollary \ref{TGIc} we have that
$$
\Big|\sum_{g\in G_n} \prod_{\{i,j\}\in E_g}\left(e^{- V_{ij}}-1\right)\Big|=
\lim_{H\to\infty }
\Big|\sum_{g\in G_n} \prod_{\{i,j\}\in E_g}\left(e^{- V^H_{ij}}-1\right)\Big|=
$$
$$
=\lim_{H\to +\infty}\Big|
\sum_{\t\in T_n}\prod_{\{i,j\}\in E_\t} (- V^H_{ij}) \int
d\m_{\t}({\bf t}_{n},{\bf X}_{n}) e^{-  \sum_{1\leq i<j\leq n}
{\bf t}_{n} (\{i,j\})V^H_{ij}}\Big|\le
$$
$$
\le \lim_{H\to +\infty}
\sum_{\t\in T_n}\prod_{\{i,j\}\in E_\t} |V^H_{ij}| \int
d\m_{\t}({\bf t}_{n},{\bf X}_{n}) e^{-  \sum_{1\leq i<j\leq n}
{\bf t}_{n} (\{i,j\})V^H_{ij}}\le
$$
$$
\le\lim_{H\to +\infty}
\sum_{\t\in T_n}\prod_{\{i,j\}\in E_\t} \Big[|\Phi^{1,H}_{ij}|+|\Phi^{2}_{ij}|\Big] \int
d\m_{\t}({\bf t}_{n},{\bf X}_{n}) e^{-  \sum_{1\leq i<j\leq n}
{\bf t}_{n} (\{i,j\})[\Phi^{1,H}_{ij}+\Phi^{2}_{ij}]}\le
$$
$$
\le e^{n B_0}\lim_{H\to +\infty}
\sum_{\t\in T_n}\prod_{\{i,j\}\in E_\t} \Big[|\Phi^{1,H}_{ij}|+|\Phi^{2}_{ij}|\Big] \int
d\m_{\t}({\bf t}_{n},{\bf X}_{n}) e^{-  \sum_{1\leq i<j\leq n}
{\bf t}_{n} (\{i,j\})\Phi^{1,H}_{ij}}
$$
where in the last line we have used the stability condition on the factor $e^{-  \sum_{1\leq i<j\leq n}
{\bf t}_{n} (\{i,j\})\Phi^{2}_{ij}}$.
Let us now define, for a fixed $\t\in T_n$ a variable $\s$ with values in $\{1,2\}$ associated to each edge of $E_\t$,
and for each edge $\{i,j\}\in E_\t$ the  numbers
$$
\Phi^\s_{ij}=\cases{|\Phi^{1,H}_{ij}| & if $\s=1$\cr\cr
|\Phi^{2}_{ij}| & if $\s=2$
}
$$
Let $\Si_\t$ be the set of all functions $\s: E_\t\to \{1,2\}$, then clearly
$$
\prod_{\{i,j\}\in E_\t} \Big[|\Phi^{1,H}_{ij}|+|\Phi^{2}_{ij}|\Big]= \sum_{\s\in \Si_\t} \prod_{\{i,j\}\in E_\t}
\Phi^{\s}_{ij}
$$
Then we can write
$$
\Big|\sum_{g\in G_n} \prod_{\{i,j\}\in E_g}\left(e^{- V_{ij}}-1\right)\Big|\le
$$
$$
\le
e^{n B_0}\lim_{H\to +\infty}
\sum_{\t\in T_n}\sum_{\s\in \Si_\t} \prod_{\{i,j\}\in E_\t}
\Phi^{\s}_{ij} \int
d\m_{\t}({\bf t}_{n},{\bf X}_{n}) e^{-  \sum_{1\leq i<j\leq n}
{\bf t}_{n} (\{i,j\})\Phi^{1,H}_{ij}}
$$
\be\label{aa}
\doteq e^{n B_0} \lim_{H\to +\infty}
\sum_{\t\in T_n}\sum_{\s\in \Si_\t} w^H_{\t,\s}
\ee
where\def\E{{\rm E}}
$$
 w^H_{\t,\s}=
\prod_{\{i,j\}\in E_\t} \Phi^{\s}_{ij}
\int d\m_{\t}({\bf t}_{n},{\bf X}_{n})
e^{- \sum_{1\leq i<j\leq n}{\bf t}_{n} (\{i,j\})\Phi^{1,H}_{ij}}
$$
Observe now that, for any tree $\t\in T_n$ and any fixed $\s\in \Si_\t$,  the edge set { $E_\t$
is naturally partitioned}  into two disjoint sets
$E_\t^1$ and $E^2_\t=E_\t\backslash E_\t^1$ where
$$
E_\t^1\,\,= \,\,\{\{i,j\}\in E_\t : \s(\{i,j\})=1\}
$$
In other words $E_\t^1$ is formed with those edges $\{i,j\}$ of $\t$ such that $\Phi^\s_{ij}=|\Phi^{1,H}_{ij}|$
and $E_\t\backslash E_\t^1$  is formed with those edges $\{i,j\}$ of $\t$ such that $\Phi^\s_{ij}=|\Phi^{2}_{ij}|$.
So
\be\label{wts}
 w^H_{\t,\s}=
\prod_{\{i,j\}\in E^1_\t} |\Phi^{1,H}_{ij}| \prod_{\{i,j\}\in E^2_\t} |\Phi^{2}_{ij}|
\int d\m_{\t}({\bf t}_{n},{\bf X}_{n})
e^{- \sum_{1\leq i<j\leq n}{\bf t}_{n} (\{i,j\})\Phi^{1,H}_{ij}}
\ee

\\The potential $\Phi^{1,H}_{ij}$ is non negative, so

$$
\sum_{1\leq i<j\leq n}
{\bf t}_{n}(\{i,j\})\Phi^{1,H}_{ij}
\ge
\sum_{\{i,j\}\in E_\t^1}
{\bf t}_{n}(\{i,j\})\Phi^{1,H}_{ij}=
$$
$$
=
\sum_{\{i,j\}\in E_\t^1}
{\bf t}_{n}(\{i,j\})\Phi^{1,H}_{ij}
+
\h\sum_{\{i,j\}\in E_\t\backslash E_\t^1}
{\bf t}_{n}(\{i,j\}) -\h
\sum_{\{i,j\}\in E_\t\backslash E_\t^1}
{\bf t}_{n}(\{i,j\})~~\ge
$$
$$
\ge
\sum_{\{i,j\}\in E_\t^1}
{\bf t}_{n}(\{i,j\})\Phi^{1,H}_{ij}+\!\!\!\!\!
\sum_{\{i,j\}\in E_\t\backslash E_\t^1}
{\bf t}_{n}(\{i,j\}))\h  - {|E_\t^2|\h}
$$
\be\label{edueb}
\ge
\sum_{\{i,j\}\in E_\t}
{\bf t}_{n}(\{i,j\})V^\t_{ij} - \eta|E_\t^2|
\ee

\\where $V^\t_{ij}$ is  the positive ($H$ dependent) pair potential given by
\be\label{defvt2}
V^\t_{ij}=\cases{\Phi^{1,H}_{ij}&if $\{i,j\}\in E_\t^1$ \cr\cr
{\h} & if $\{i,j\}\in  E_\t^2$}
\ee

\\Plugging (\ref{edueb}) into   (\ref{wts}), we can write

$$
w^H_{\t,\s}\le
\left[{e^\h\over \h}\right]^{|E_\t^2|}
\prod_{\{i,j\}\in E^2_\t} |\Phi^{2}_{ij}|
{ \prod_{\{i,j\}\in  E_\t}  V^\t_{ij}
 \int d\m_{\t}({\bf t}_{n},{\bf X}_{n})
e^{-  \sum_{\{i,j\}\in E_\t}
{\bf t}_{n}(\{i,j\})V^\t_{ij} }}
$$

\\Using now Lemma \ref{posit}, i.e. formula (\ref{TGI3}), we have that
$$
{\prod_{\{i,j\}\in  E_\t}  V^\t_{ij}
 \int d\m_{\t}({\bf t}_{n},{\bf X}_{n})
e^{-  \sum_{\{i,j\}\subset E_\t}
{\bf t}_{n}(\{i,j\})V^\t_{ij} }}=
\prod_{\{i,j\}\in E_\t}\left|e^{-V^\t_{ij}}-1\right|=
$$
$$
=
 \prod_{\{i,j\}\in E_\t^1}\left|e^{-\Phi^{1,H}_{ij}}-1\right|
\left|e^{-\h}-1\right|^{|E_\t^2|}
$$
Thus we get,

$$
w^H_{\t,\s}\le
\left[\prod_{\{i,j\}\in E_\t^2} {(e^\h-1)\over \h}|\Phi^{2}_{ij}|\right]
 \prod_{\{i,j\}\in E_\t^1}\left|e^{-\Phi^{1,H}_{ij}}-1\right|=
$$

\\Hence, since $\h$  can be taken as small as we please, we get finally

\be\label{whts}
w^H_{\t,\s}\le
\left[\prod_{\{i,j\}\in E_\t^2} |\Phi^{2}_{ij}|\right]
 \prod_{\{i,j\}\in E_\t^1}\left|e^{-\Phi^{1,H}_{ij}}-1\right|
\ee
Plugging (\ref{whts}) into (\ref{aa}) we get
$$
\Big|\sum_{g\in G_n} \prod_{\{i,j\}\in E_g}\left(e^{- V_{ij}}-1\right)\Big|\le
$$
$$
\le
e^{n B_0}\lim_{H\to +\infty}
\sum_{\t\in T_n}\sum_{\s\in \Si_\t} \left[\prod_{\{i,j\}\in E_\t^2} |\Phi^{2}_{ij}|\right]
 \prod_{\{i,j\}\in E_\t^1}\left|e^{-\Phi^{1,H}_{ij}}-1\right|=
$$
$$
=
e^{n B_0}
\sum_{\t\in T_n}\sum_{\s\in \Si_\t} \left[\prod_{\{i,j\}\in E_\t^2} |\Phi^{2}_{ij}|\right]
 \prod_{\{i,j\}\in E_\t^1}\left|e^{-\Phi^{1}_{ij}}-1\right|=
$$
$$
=
e^{n B_0}
\sum_{\t\in T_n} \prod_{\{i,j\}\in E_\t} \Big[|\Phi^{2}_{ij}|+ \left|e^{-\Phi^{1}_{ij}}-1\right| \Big]
$$
$\Box$

\section{Proof of Theorems \ref{teo1} and \ref{teo2}}
\zeq
We can now use Theorems \ref{stabbry} and \ref{stabru} straightforwardly to obtain the new bounds for the Mayer coefficient and consequently the Mayer series radius of convergence
as far as Penrose potentials and Ruelle Potentials are concerned.

\subsection{\bf Proof of Theorem \ref{teo1}}
Let $V(|x|)$ be Penrose pair potential according to Definition \ref{hard2a} and let $B$ be its stability constant. Then, for any $n\in \mathbb{N}$ and $(x_1,\dots,x_n)\in \mathbb{R}^{dn}$,
$V_{ij}=\b V(|x_i-x_j|)$ (with $1\le i<j\le n$) is a stable algebraic potential in $[n]$ with stability constant $\b B$ , according to definitions \ref{pair} and \ref{staby}. Therefore we can use Theorem \ref{stabbry} to get a bound on the $n$-order coefficient $C_n(\b,\L)$, defined in (\ref{urse}), of the Mayer series of the system of particles interacting via the pair potential $V(|x|)$. The bound goes as follows.
$$
|C_n(\b,\L)|\le  {1\over |\L| }{1\over n!}\int_{\L}dx_1 \dots \int_{\L}
dx_n \left|\sum\limits_{g\in G_{n}} \prod\limits_{\{i,j\}\in E_g}\left[
e^{ -\b V(|x_i -x_j|)} -1\right]\right| \le
$$
\be\label{pteo1}
\le  {1\over |\L| }{1\over n!} e^{n  \b B} \sum_{\t\in T_n} \int_{\L}dx_1 \dots \int_{\L}
dx_n \prod_{\{i,j\}\in E_\t}
F(|x_i-x_j|)
\ee
where
$$
F_\b(|x_i-x_j|)= \cases{ |e^{- \b V(|x_i-x_j|)}-1|\equiv 1 & if $V(|x_i-x_j|)=+\infty$ \cr\cr
 \b |V(|x_i-x_j)| & otherwise
}
$$
Let now  $a$ be  the hard-core radius of the potential $V(|x|)$. Then, recalling  (\ref{cstarb}), we have
$$
\int_{\mathbb{R}^d} F(|x|)=   \int_{|x|\le a} 1\; dx + \b\int_{|x|\ge a} |V(|x|)| dx= C^*(\b)
$$
Moreover, using  standard observations in cluster expansions, we have  that, for any tree $\t\in T_n$
\be\label{clus}
\int_{\L}dx_1 \dots \int_{\L}
dx_n \prod_{\{i,j\}\in E_\t} F(|x_i-x_j|) \le |\L| \Big(\int_{\mathbb{R}^d} F(|x|) dx\Big)^{n-1}= |\L| \Big[C^*(\b)\Big]^{n-1}
\ee
Plugging now (\ref{clus}) into (\ref{pteo1}) and recalling that $\sum_{\t\in T_n} 1= n^{n-2}$, we get the desired bound (\ref{bo1}). $\Box$

\subsection{\bf Proof of Theorem \ref{teo2}} Proceeding similarly as above,
let now $V(|x|)=\Phi_1(|x|)+\Phi_2(|x|)$ denote a Ruelle pair potential according to Definition \ref{ruelle} and let $\tilde B$ be the stability constant of the potential $\Phi_2(|x|)$. Then, for any $n\in \mathbb{N}$ and $(x_1,\dots,x_n)\in \mathbb{R}^{dn}$,
$V_{ij}=\b V(|x_i-x_j|)$ (with $1\le i<j\le n$) is a Ruelle-stable algebraic potential in $[n]$ according to Definition \ref{staru}, such that
$$
V_{ij}= \Phi^1_{ij}+{\Phi^2}_{ij}~~~~~~{\rm  with}~~~~~~  \Phi^1_{ij}=\b \Phi_1(|x_i-x_j|)~~~~ {\rm and}~~~~ \Phi^2_{ij}=\b \Phi_2(|x_i-x_j|)
$$
Moreover  ${\Phi^2}$ has  stability constant $\b \tilde B$.
We can now  use Theorem \ref{stabru} to  bound the $n$-order Mayer coefficient as follows.
$$
|C_n(\b,\L)|\le  {1\over |\L| }{1\over n!}\int_{\L}dx_1 \dots \int_{\L}
dx_n \left|\sum\limits_{g\in G_{n}} \prod\limits_{\{i,j\}\in E_g}\left[
e^{ -\b V(|x_i -x_j|)} -1\right]\right| \le
$$
\be\label{pteo2}
\le  {1\over |\L| }{1\over n!} e^{n  \b \tilde B} \sum_{\t\in T_n}\int_{\L}dx_1 \dots \int_{\L}
dx_n
\prod_{\{i,j\}\in E_\t}
\Big[|e^{-\b \Phi_1(|x_i-x_j|)}-1|+ |\b \Phi_2(|x_i-x_j|)|\Big]
\ee
Recalling  now (\ref{tilb}) and proceeding in a completely analogous manner as we did in the proof of Theorem \ref{teo1}, we have  that, for any tree $\t\in T_n$
$$
\int_{\L}dx_1 \dots \int_{\L}
dx_n \prod_{\{i,j\}\in E_\t} \Big[|e^{-\b \Phi_1(|x_i-x_j|)}-1|+ |\b \Phi_2(|x_i-x_j|)|\Big] \le
$$
\be\label{clus2}
\le  |\L| \Big(\int_{\mathbb{R}^d}\Big[|e^{-\b \Phi_1(|x|)}-1|+ |\b \Phi_2(|x|)|\Big] ) dx\Big)^{n-1}= |\L| \Big[\tilde C(\b)\Big]^{n-1}
\ee
Plugging now (\ref{clus2}) into (\ref{pteo2}) and using again  that $\sum_{\t\in T_n} 1= n^{n-2}$, we get the  bound (\ref{bteo2}). $\Box$

\renewcommand{\theequation}{A.\arabic{equation}}
\appendix
\section*{Appendix: proof of Proposition \ref{lj}}
\zeq

\\Let $V(|x|)$ be a Lennard-Jones type potential according to Definition \ref{lj}, let  $a>0$ and let
\be\label{Va}
V_a(|x|)= \cases{ V(|x|) & if $|x|\ge a$\cr\cr
V(a) & if $|x|<a$
}
\ee
We will first prove  that, for sufficiently small $a\in (0,r_1)$,  the potential $V_a(|x|)$ defined in (\ref{Va}) is stable by showing that it can be written as a sum of a positive  potential plus a positive-definite potential  (see \cite{Ru, ga} and see also remark after definition \ref{ruelle}).
We
basically follow the strategy adopted by  Fisher and Ruelle  \cite{FR} who showed the same in case of Lennard-Jones type potentials.  The proof is developed in three steps.
\vv
\\1. We first construct a bounded monotonic decreasing tempered and non-negative function $\eta_3(|x|)$ such that
\be\label{uno}
V_a(|x|)\ge -\eta_3(|x|)
\ee
and such that the Fourier transform $\hat{\eta}_3(p)$ of  $\eta_3(r)$ (which exists since $\eta_3$ is absolutely integrable) is bounded as
\be\label{uno2}
|\hat{\eta}_3(p)| \le {C_1\over( |ap|^2 +1)^d}
\ee
where $C_1$ is a constant and $a$ is the same constant appearing in (\ref{Va}). Recall that, since $\eta_3(|x|)$ is a
radial function in $\mathbb{R}^d$, then its Fourier transform is
also radial and moreover $\hat{\eta}_3(p)$ is real.
\vv
\\2.   We then construct a bounded monotonic compact support (hence tempered) non negative function $\xi_1(|x|)$ such that
\be\label{due}
\xi_1(|x|)~~\cases{ \le V_a(|x|) & if $|x|\le r_1$\cr\cr
0 & otherwise
}
\ee
and  the Fourier transform $\hat{\xi}_1(p)$ of  $\xi_1(|x|)$ (which exists since $\xi_1(|x|)$ is absolutely integrable) is positive and bounded as
\be\label{due2}
\hat{\xi}_1(p)\ge  {C_2\over( |ap|^2 +1)^d}
\ee
with $C_2$  constant and $a$ is the same constant appearing in (\ref{Va}).
\vv
\\3. Once functions $\eta_3(|x|)$ and $\xi_1(|x|)$ with the properties (\ref{uno}), (\ref{uno2}), (\ref{due}), (\ref{due2})  have been constructed we write
$$
V_a(|x|)=  \Psi_1(|x|) +\Psi_2(|x|)
$$
with
$$
\Psi_1(|x|)= V_a(|x|) + \eta_3(|x|)-\xi_1(|x|)
$$
and
$$
\Psi_2(|x|)=\xi_1(|x|)-\eta_3(|x|)
$$
Note that  both $\Psi_1(|x|)$ and $\Psi_2(|x|)$ are tempered since they are sum of tempered functions (recall that $V(|x|)$ is assumed tempered by hypothesis).
Moreover,  by (\ref{uno}) and (\ref{due}) we have immediately that $\Psi_1(|x|)\ge 0$ for all $|x|\ge 0$.
Finally, by (\ref{uno2})  and (\ref{due2})  we have that  $\Psi_2(|x|)$ has non negative fourier transform and thus is positive definite.
\vv\vv
\\{\it 1. Construction of the function  $\eta_3(|x|)$}.
Let us  construct the function $\eta_3(r)$ with properties (\ref{uno}) and (\ref{uno2}). We first pose
$$
\eta_1(|x|)=\cases{ w & if $|x|\le r_2$ \cr\cr
\eta(|x|) & if $|x|> r_2$
}
$$
So that, for all $|x|\ge 0$, we have that
\be\label{st1}
V_a(|x|)\ge - \eta_1(|x|)
\ee
Then we, letting $a\in (0,r_1)$,  define
\be\label{stt1}
\eta_2(|x|)=\cases{ w & if $|x|\le a$ \cr\cr
\eta_1(|x|-a) & if $|x|\ge a$
}
\ee
Of course, by construction,  $\eta_1(|x|)$ and $\eta_2(|x|)$ are absolutely summable functions  (since $\eta(|x|)$ is tempered).
Indeed, if we let $H= \int_{|x|\ge r_2}\eta(|x|) dx$. Then
$$
\|\eta_1\|_1~= ~{\pi^{d\over 2}r_2^d\over \Gamma({d\over 2}+1)}w+ H
$$
and
$$
\|\eta_2\|_1~=~ {\pi^{d\over 2}(r_2+a)^d\over \Gamma({d\over 2}+1)} w + H~ \le ~{\pi^{d\over 2}(r_2+r_1)^d\over \Gamma({d\over 2}+1)} w + H
$$
Moreover we have that
\be\label{st2}
\eta_2(|x'|)\ge \eta_1(|x|)~~~~~~~~~~~~~{\rm if}~~  ||x|-|x'||\le a
\ee
Now take a  non-negative function $\psi(|x|)$ such that $\psi(|x|)=0$ whenever $|x|>1$ and such that
\be\label{st3}
\int_{\mathbb{R}^d} \psi(|x|)dx =1
\ee
It is always possible to choose this function $\psi$ in such a way that it has continuous derivatives of all orders in such a way that its Fourier transform decays at large
distances faster than any inverse polynomial. In other words it is always possible to find a (universal) constant $C'$
such that, if $\tilde \psi(p)=\int \psi(|x|)e^{ip\cdot x}d x$ is the Fourier transform of $\psi(|x|)$, we have that
$$
|\tilde\psi(p)|\le {C'\over (1+ p^2)^d}
$$
Note that $\tilde\psi(p)$ is real and radial.
Let now
$$
\psi_a(|x|)=
{1\over a^d}~ \psi\Big({|x|\over a}\Big)
$$
We have clearly that
\be\label{st32}
\int_{\mathbb{R}^d} \psi_a(|x|)dx =1
\ee
and $\psi_a(|x|)=0$ if $|x|> a$. Define now
$$
\eta_3(|x|)=\int \psi_a(|x-x'|)\eta_2(x') dx'
$$
Due to(\ref{st2})  and to the fact that $\psi_a(|x|)=0$ if $|x|>a$, we have that
$$
\eta_3(|x|)=\int \psi_a(|x-x'|)\eta_2(|x'|) dx'\ge \int \psi_a(|x-x'|)\eta_1(|x|)dx' = \eta_1(|x|)
$$
and hence, due to (\ref{st1}) we have that
$$
V_a(|x|)\ge -\eta_3(|x|)
$$
Moreover $\eta_3(|x|)$ is absolutely integrable (since, by (\ref{st3})) ,  $\|\eta_3\|_1= \|\eta_2\|_1$) and its Fourier transform is
$$
\tilde \eta_3(p)= \tilde \eta_2(p) \tilde\psi_a(p)= \tilde \eta_2(p) \tilde\psi(ap)
$$
So that
$$
|\tilde \eta_3(p)|\le\|\eta_2\|_1 |\tilde\psi(ap)|\le {\left[{\pi^{d\over 2}(r_2+r_1)^d\over \Gamma({d\over 2}+1)} w + H \right]C'\over [1+ (ap^2)]^d}
$$
\vv\vv

\\{\it 2. Construction of the function  $\xi_1(r)$}. Take a radial function $\chi(|x|)$ (not identically zero) which is non negative, continuous and vanishes for $|x|\ge {1\over 2}$.
Let now
$$
\chi_1(|x|)= \int dx'\chi(|x-x'|)\chi(|x'|)
$$
By construction $\chi_1(|x|)$ is continuous non negative and vanishes for $|x|\ge 1$. Moreover the Fourier transform $\tilde\chi_1(p)$ of $\chi_1(|x|)$
is non negative (since it is the square
of the Fourier transform of $\chi(|x|)$ which is radial and real) and it is non zero in some neighbor of $p=0$ (since $\tilde\chi_1(p=0)>0)$.
Consider now
\be\label{st4}
\chi_2(|x|)= \chi_1(|x|)\Psi(|x|)
\ee
where
$$
\Psi(|x|)=\int dp {e^{ip\cdot x} \over (p^2 +1)^d}
$$
Then $\chi_2(|x|)$ is non negative  (since the integral in (\ref{st4}) is  the modified Bessel function of the third kind) and  vanishes for $r\ge 1$.
Moreover its Fourier transform is
$$
\tilde\chi_2(p)=\int dp' {\tilde\chi_1(p') \over [(p-p')^2 +1]^d}\ge {C''\over  [p^2 +1]^d}
$$
where in the last line we have used the fact that $\tilde\chi_1(p')$ is  strictly positive at  $p=0$ and it is continuous.
Let now  $K =\max \{\chi_2(x)\}$ and define
$$
\chi_3(|x|) = {1\over K} \chi_2(|x|)
$$
Then $\chi_3(|x|) $  vanishes for $|x|\ge 1$, and  $ \chi_3(|x|)\le 1$ when  $|x|< 1$. Moreover its Fourier transform $\tilde\chi_3(p)$ is such that
$$
\tilde\chi_3(p)\ge {C^*\over   [p^2 +1]^d}
$$
where
$$
C^*={C''\over K}
$$
Let now, for  $a\in (0,r_1)$  previously introduced,
$$
\theta_a(r) =\cases{ 1 & if $r\le a$ \cr\cr 0 &if  $r>a$
}
$$
Then  the function
$$
\phi(|x|)= \xi(a)\theta_{a} (|x|)
$$
is by construction such that
$$
\phi(|x|)\le V_a(|x|)~~~~~~{\rm for ~all}~~|x|\le r_1
$$
Let finally pose
$$\xi_1(|x|)= \xi(a)~ \chi_3({|x|\over a})$$
Since $\chi_3({r\over a})\le \theta_{a} (r)$, we have that
$ \xi_1(r)\le \phi(r)\le \xi(r)$, for all $|x|\le r_1$.  Moreover
$$
\tilde\xi_1(p) =a^d\xi(a)\tilde \chi_3(ap)\ge   { C^* }{a^d\xi(a)\over [(a p)^2+1]^d}
$$

\\We can now bound the Fourier transform of $\Phi_2(|x|)= \xi_1(|x|) - \eta_3(|x|)$ as

\be\label{Four}
\tilde \Phi_2(p)\ge
\left[{C^* a^d\xi(a) }- C' \Big[{\pi^{d\over 2}(r_2+r_1)^d\over \Gamma({d\over 2}+1)}w + H \Big]\right]
{1\over [(a p)^2+1]^d}
\ee
We may now choose  our $a\in (0,r_1)$ such that
\be\label{xC}
\x(a)\ge{ C\over a^d}
\ee
with
$$
C ={C'\over C^*} {\left[{\pi^{d\over 2}(r_2+r_1)^d\over \Gamma({d\over 2}+1)} w + H \right]}
$$
which is always possible, since, due to the assumption (\ref{short}), $\x(|x|)|x|^d\to \infty$ as $|x|\to 0$.
By inserting condition (\ref{xC}) into (\ref{Four}) we get  that
$$
\tilde \Phi_2(p)\ge 0
$$
I.e. $\Phi_2(|x|)$ has a non-negative Fourier transform and so it  is positive-definite. Hence, when $a$ satisfies (\ref{xC}), $V_a(|x|)= \Psi_1(|x|) +\Psi_2(|x|)$,
being  the sum  of a positive $\Psi_1(|x|)$
plus a positive-definite  $\Psi_2(|x|)$  is, by the Ruelle criterium \cite{Ru}, a stable pair potential.

\vv
\\We can now conclude the proof of Proposition \ref{lj}.  We have proved above that $V_a(|x|)$ is stable. We note now that $V_a(|x|)$ is, by construction, absolutely summable
(because $V_a(|x|)$ is bounded by $V(a)$ at short distances and moreover $V(|x|)$ is tempered). Hence we can write
$$
V(|x|)= \Phi_1(|x|)+ \Phi_2(|x|)
$$
where
$\Phi_1(|x|)= V(|x|) - V_a(|x|)$ non-negative and tempered (actually compact supported)  and $\Phi_2(|x|) =V_a(|x|)$ stable and absolutely summable.

\\$\Box$

\section*{Acknowledgments}
A. P. has   been partially supported by the Brazilian  agencies
Conselho Nacional de Desenvolvimento Cient\'{\i}fico e Tecnol\'ogico
(CNPq) and FAPEMIG
(Funda{c}\~ao de Amparo \`a Pesquisa do Estado de Minas
Gerais)-Programa Pesquisador Mineiro.

\end{document}